# ·Zero-field Anomalous Hall Effect in Bulk Single Crystal Mn$_3$Ir

*Xin Gu[1], Ruoqi Wang[1], Bo Zhao[2], Haofu Wen[1], Kunquan Hong[1], Shijun Yuan[1]\*, Taishi Chen[1]\*, Jinlan Wang[1]*

1. Key Laboratory of Quantum Materials and Devices of Ministry of Educations, School of Physics, Southeast University, Nanjing 211189, China
2. Shanghai Advanced Research Institute, CAS 239 Zhangheng Road, Pudong, Shanghai 201204, China.

## Abstract:

The L1$_2$-phase non-collinear antiferromagnet (AFM) Mn$_3$Ir has emerged as a pioneering platform for realizing the zero-field giant anomalous Hall effect (AHE), thereby catalyzing the rapid advancement of antiferromagnetic spintronics. Despite its significant potential, the experimental investigation of the intrinsic magnetic and electronic properties of Mn$_3$Ir has been greatly hindered, primarily due to the formidable challenges associated with the growth of bulk single crystals. Here, we report the successful growth of stoichiometric Mn$_3$Ir bulk single crystals, and characterization on the magnetization and AHE. We successfully obtained (111)-oriented hexagonal Mn$_3$Ir single crystals using a high-throughput flux method. The smaller AHE was successfully detected. We attribute intriguing results to the coexistence of A-phase and B-phase antiferromagnetic domains mutually cancelling the AHE response. Our work reveals the key details of the intrinsic magnetic properties and AHE in bulk Mn$_3$Ir. This provides a key material for the development of advanced antiferromagnetic spintronic devices.

Keywords: Mn$_3$Ir, bulk single crystal, Kagome antiferromagnet, anomalous Hall effect, noncollinear antiferromagnets

Xin Gu, Ruoqi Wang and Bo Zhao contributed equally to this work; Corresponding author: chentaishi@seu.edu.cn, siesta@seu.edu.cn

# 1. Introduction

The study of the AHE in magnetic materials is a central topic in condensed matter physics, spanning interdisciplinary fields such as topological electronics, thermodynamics, photonics, and information storage, etc.[1-6] Particularly, it holds profound implications for the advancement of high-performance spintronic memory devices and quantum computing. The Berry curvature-based band theory has successfully explained a range of enhanced transverse transport phenomena observed in ferromagnets,[1, 2, 7-9] including the experimental realization of the quantum anomalous Hall effect.[4, 10] However, the inherently strong net magnetization in ferromagnetic materials presents three major technical bottlenecks:[11, 12] (1) magnetic crosstalk between neighboring storage units limits storage density; (2) the high energy cost of domain reversal hinders low-power operation; and (3) restricted magnetodynamic response speeds constrain high-speed read/write performance. Therefore, the development of novel material systems that simultaneously exhibit tiny net magnetization and a strong AHE is key to breakthroughs in next-generation ultrahigh-density, ultrahigh-speed, and ultralow-power magnetic storage technologies.

In 2014, Chen *et al.* theoretically predicted that $L1_2$-phase AFM $Mn_3Ir$ single crystal can produce an intrinsic AHE as large as 218 S/cm, only driven by a combination of the triangle antiferromagnetic structure in Kagome lattice and strong spin-orbit coupling.[13] The establishment of this theory immediately catalyzed rapid progress in antiferromagnetic research,[14-21] and led to the emergence of a new field: antiferromagnetic spintronics.[11, 12] However, as the foundational material for this theoretical framework, $Mn_3Ir$ bulk single crystal has long eluded successful synthesis due to the complexity of the Mn–Ir binary phase diagram and the extremely low solubility of Ir atoms in high-temperature melts.[22]

In this study, we report the successful growth of oriented centimeter-scale $Mn_3Ir$ bulk single crystals via an innovatively designed high-throughput flux method. Basic characterizations confirm that the crystals are high-quality, stoichiometric (Mn:Ir = 3:1) single crystals, with Mn spins arranged in the (111) non-collinear antiferromagnetic structure as previously reported.[23] Crucially, I found that bulk single-crystal Mn$_3$Ir exhibits distinct behavior in both its magnetic properties and AHE, setting it apart from previous experimental reports.[24-30] Notably, bulk Mn$_3$Ir shows weak ferromagnetic saturation magnetization of only ~0.1 m$\mu_B$ per Mn and switches easy magnetization axes at low and high temperatures. Furthermore, we successfully detected the AHE in bulk single crystal in both low and high temperature, though its magnitude remains as low as 2 S/cm. We attribute this anomaly to the coexistence of A- and B-phase antiferromagnetic domains in the sample (as defined by Hua Chen *et al.*), which leads to mutual cancellation of Berry curvature contributions.[13] Considering the recent rapid advances in methods for manipulating antiferromagnets, we believe that a wide range of spintronic effects can be observed in bulk Mn$_3$Ir.[26, 31-37]

## 2. Main text

### 2.1. Evaluation of Preferred Orientation Growth of Mn$_3$Ir Single Crystals

**Figure 1** presents the cleavage energy of Mn$_3$Ir along various crystallographic planes under different magnetic configurations, calculated via density functional theory (DFT). Prior studies have established Mn$_3$Ir adopts the L1$_2$-type ordered FCC structure, where Ir atoms are located at the cube corners and Mn atoms are situated at the face centers of the cubic lattice.[38] Crucially, the Mn magnetic moments form an A or B-phase non-collinear antiferromagnetic arrangement confined to the Kagome (111) plane as was previously defined.[23] This distinctive magnetic structure typically can induce large magnetocrystalline anisotropy, profoundly influencing preferential crystal growth orientations.

In this work, we investigate two magnetic configurations (as shown in Figure **1a**): (i) the non-collinear triangular antiferromagnetic order reported in previous literature, and

(ii) a ferromagnetic state with all spins aligned along the [001] direction. Following full structural and relaxation, we computed the cleavage energies of Mn$_3$Ir along the <001>, <110>, and <111> directions, as summarized in Figure **1b**. Our results reveal that the antiferromagnetic phase exhibits a pronounced minimum cleavage energy on the (111) plane, with a significant energy difference of approximately 0.02 eV/Å$^2$ compared to the other two orientations. In contrast, differences in cleavage energy for the ferromagnet (FM) state are negligible across these orientations, although the (111) plane also exhibits the lowest value. This marked anisotropy in the antiferromagnetic state strongly suggests that Mn$_3$Ir crystals grown in the L1$_2$ (FCC) phase will preferentially form hexagonal morphologies, potentially even layered hexagonal flakes.[39, 40] From the standpoint of probing antiferromagnetic magnetism and the associated AHE, achieving growth with the non-collinear antiferromagnetic order on the (111) plane constitutes our primary experimental objective.[13]

## 2.2. Experimental growth and structural characterization of Mn$_3$Ir

**Figure 2** illustrates the experimental growth for the bulk single-crystal of Mn$_3$Ir and its structural characterization. Based on a comprehensive survey of binary phase diagrams for Mn–Ir, Mn–Bi, and Ir–Bi systems, we selected bismuth (Bi) as a flux agent to facilitate Mn$_3$Ir crystal growth.[22, 41, 42] Considering the extremely low solubility of Ir in Bi, we developed a high-throughput flux method to efficiently optimize the composition ratios of Bi, Mn, and Ir. To ensure sufficient Ir content in the melt, we employed the sawtooth-shaped cooling profile depicted in Figure **2a**. After approximately 50 batches of trials, we successfully obtained Mn$_3$Ir single-crystal platelets, as shown in the inset of Figure **2a**, some reaching dimensions up to 1.5 cm × 1.5 cm × 0.05 mm.

The X-ray diffraction (XRD) patterns of the resulting samples are also presented in Figure **2b**. It is worth noting that Mn$_3$Ir crystals are relatively soft and challenging to grind into powder, so the XRD data were collected from multiple single-crystal platelets.

Due to the preferred orientation of crystal growth, the diffraction peaks corresponding to the (111) family are prominently strong, while other peaks appear much weaker. Comparison with standard crystallographic databases confirms that the crystals possess the $L1_2$ face-centered cubic structure. The inset in Figure **2b** further reveals additional diffraction peaks characteristic of the $L1_2$ phase. Residual elemental Bi peaks, attributed to incomplete removal of flux particles adhering to the $Mn_3Ir$ surface, are also visible.

To unambiguously determine the crystallographic orientation, we performed Laue single-crystal diffraction using a BR001 Laue system (Photonic Science) equipped with a high-precision goniometer (also seen in Supplementary Figure **S2**). When the X-rays were incident perpendicular to the sample surface, sharp diffraction spots consistent with the simulated Laue pattern for the $L1_2$ (111) plane were collected, as shown in Figure **2c-2e**. By rotating the sample, we successfully observed well-defined Laue spot arrays corresponding to the (110) and (100) planes at tilt angles of 35.264° and 54.736° relative to [111], respectively.

In combination with the Scanning Electron Microscopy (SEM) / Energy Dispersive Spectroscopy (EDS) results yielding the perfect atomic ratio Mn:Ir of 3:1 (Details shown in Supplementary Figure **S1**), we have realized the growth of high-quality $Mn_3Ir$ single-crystal platelets exhibiting the $L1_2$ structure with [111] out-of-plane orientation. Subsequently, we proceed to characterize the magnetic properties and anomalous Hall effect of these $Mn_3Ir$ single crystals.

### 2.3. Magnetization of bulk $Mn_3Ir$ single crystal

**Figure 3** show the magnetic characterization on the bulk single crystal $Mn_3Ir$. In Figure **3a**, the out-of-plane (111 direction) magnetization at various temperatures is exhibited. Non-collinear antiferromagnetic structures with Kagome lattice geometry play a central role in recent antiferromagnetic spintronics, mainly because such structures host a variety of nontrivial magnetic orders breaking time-reversal symmetry, thereby generating nonzero Berry curvature and AHE. In realistic antiferromagnets, however, the presence of nearly degenerate antiferromagnetic domains—with almost equal populations and opposite orientations—tends to significantly cancelling intrinsic transverse responses. This underpins the focus on antiferromagnets exhibiting weak residual (net) magnetization due to canting, as they provide a crucial experimental handle for the control of magnetic domains and the direction of time-reversal symmetry breaking.

As shown in Figure **3a**, the out-of-plane magnetization is of the order of 0.1 m$\mu_B$/ Mn, and increases slowly and linearly with the applied magnetic field. Notably, in the low-field regime (<5000 Oe), the magnetization shows a clear hysteresis loop, reaching a maximum at 300 K. This behavior is consistent with previous reports of a weakly canted magnetic structure in Mn$_3$Ir along the [111] direction, although the measured magnitude here is smaller.[13, 40, 43] This subtle net magnetization is particularly important, as it enables effective manipulation of AFM domains and the time-reversal symmetry breaking direction.[13]

Figure **3b** displays the in-plane (within the (111) plane) magnetization curves at different temperatures, with the inset showing the MH curves at low fields. It can be seen that the in-plane magnetization is an order of magnitude weaker than the out-of-plane component. At 2 K, clear weak ferromagnetic features emerge, with the extrapolated residual magnetization yielding 0.063 $m\mu_B$/ Mn, and reaches the 0.186 $m\mu_B$/ Mn at 300 K. By comparing both the in-plane and out-of-plane data (see also Figure **3c**), we find that Mn$_3$Ir exhibits distinct easy axis behaviors in different orientations depending on the temperature and this tuning point is at around 150 K, which has not been reported previously.

Figure **3d** shows the temperature dependence of the magnetization (MT) measured under a 1.5 T magnetic field. Given the extremely high Néel temperature and the minute net ferromagnetic moment of Mn$_3$Ir, it is challenging to detect such subtle temperature-

dependent changes in the net saturation magnetization. However, there is a kink point happened at 150 K for MT under $\mu_0 H \perp [111]$ direction. This greatly strengthens our confidence in investigating the AHE in Mn₃Ir bulk single crystal.

## 2.4. Magneto-transport results of Mn₃Ir single crystals

**Figure 4** presents the magneto-transport results of Mn₃Ir single crystals for both (111) in-plane and out-of-plane configurations, as illustrated schematically in Figures **4a and 4d**. Following the protocol recommended by Hua Chen et al., in which a strong magnetic field is applied along the [111] direction at high temperature to promote a single magnetic domain state (A or B), we cooled the samples and subsequently performed measurements. Using this approach, we obtained both in-plane and out-of-plane resistivity–temperature (RT) curves (Figures **4a and 4d**), Hall resistivity (Figures **4b and 4e**), and longitudinal magnetoresistance (MR) (Figure **4f**). A pronounced resistivity anisotropy between the (111) in-plane and out-of-plane orientations is clearly observed, which directly reflects the large magnetocrystalline anisotropy intrinsic to Mn₃Ir.

We next focus on AHE in Mn₃Ir bulk single crystal. Figures **4b, 4c, and 4e** show the AHE for both (111) in-plane and out-of-plane configurations at various temperatures. The insets display the Hall resistivity after subtraction of the high-field background. In Figures **4b and 4c**, we compare the AHE following field-cooling and zero-field-cooling procedures, respectively. For clarity, we define the coordinate axes as: the [111] direction as the *z* direction, the in-plane direction along the vertex of the hexagon as *y*, and the in-plane direction perpendicular to the hexagon edge as *x* (see inset of Figure **4a**).

Remarkably, the evolution of the anomalous Hall resistivity closely mirrors that of the magnetization: at low temperatures, $\rho_{xy}$ is nearly zero while $\rho_{xz}$ reaches its maximum; the trend reverses at higher temperatures. Quantitatively, both $\rho_{xy}$ and $\rho_{xz}$ are small, with values of ~0.2 $\mu\Omega\cdot$cm and ~0 $\mu\Omega\cdot$cm at 300 K, ~0.015 $\mu\Omega\cdot$cm and ~0.2 $\mu\Omega\cdot$cm at 2 K, respectively. According to the formula $\sigma_{xy} = -\rho_{xy}/(\rho_{xx}^2+\rho_{xy}^2)$, the anomalous Hall conductivities $\sigma_{xy}$ and $\sigma_{xz}$ are determined to be approximately ~2.0 S/cm and ~0 S/cm at 300 K, ~0.5 S/cm and ~2.4 S/cm at 2 K, respectively—significantly lower than the theoretical predictions of Hua Chen et al.[13] At this point, it should be emphasized that: (1) field cooling and zero-field cooling of the sample can indeed have a significant impact on AHE; (2) furthermore, we subsequently observe a sharp switch in the magnitude of the AHE at 150 K. Although the exact change in magnetic structure at 150 K remains unclear, it can be inferred that this transition leads to a reorientation of the weak ferromagnetic easy magnetization axis in Mn$_3$Ir.

## 3. Conclusion and Discussion

The observation of such a small yet finite AHE is particularly intriguing. Firstly, our Mn$_3$Ir bulk crystals are of exceptionally high quality, as supported by previous neutron diffraction studies of L1$_2$-phase Mn$_3$Ir with a non-collinear antiferromagnetic structure,[23] alongside our own comprehensive structural and compositional characterizations. Secondly, the marked magnetic anisotropy of the magnetization curves at different temperatures rules out the possibility that residual ferromagnetic Mn impurities contribute to the AHE. Moreover, the relatively low carrier mobility in both in-plane and out-of-plane directions excludes extrinsic effects such as skew scattering and side-jump mechanisms. Therefore, the anomalous Hall conductivity (AHC) observed in Mn$_3$Ir single crystals is most likely attributable to the intrinsic scattering of Bloch electrons.

Chen Hua et al. pointed out that the A and B magnetic domains in an external magnetic field possess different energies, with a difference of approximately 1 meV. Although this energy difference is comparable to the ambient temperature, it nonetheless leads to an imbalance in the populations of these two phases of domains, thereby inducing a net AHE. Figure **5** shows the temperature dependence of the AHC for both in-plane and out-of-plane (111) orientations. Although the signals are weak (2 and 2.4 S/cm), they nonetheless indicate that this non- collinear antiferromagnetic structure plays a crucial role in the intrinsic anomalous Hall effect. This anomalous Hall effect originates from the Berry curvature, which is further shown by the scaling relationship between anomalous Hall conductivity and magnetization, as shown in Supplementary Figure **S3**.

Recent advances in antiferromagnetic research further suggest strategies to enhance the AHE in $Mn_3Ir$ crystals by increasing the population imbalance between A and B domains. Such strategies may include crystal growth under applied magnetic fields or the application of external strain to the bulk $Mn_3Ir$ crystal.

In summary, our experimental progress holds significant implications both for uncovering the intrinsic mechanisms of AHE and for advancing the development of high-temperature antiferromagnetic spintronic devices. In fact, we have achieved important progress in this direction, as demonstrated in Figure **S4** of the supplementary materials.

## 4. Methods

We successfully grew high-quality bulk single crystals of $Mn_3Ir$ using a high-throughput flux method. The so-called high-throughput flux method refers to selecting multiple compositions across a multinary alloy phase diagram (in this study, more than fifty different points selected), enabling the identification of favorable conditions for target crystal growth in a single batch synthesis. We prepared fifty samples with different Bi:Mn:Ir ratios, loaded them into fifty high-purity ceramic crucibles, and then sealed the crucibles into ten quartz tubes (five crucibles per tube). Each tube was purged with high-purity argon five times, evacuated to 1 Pa, and then flame-sealed. The sealed quartz tubes were placed in two box furnaces. The temperature program was as follows: the furnace was ramped to 1050 °C within 60 minutes and held for 3 days, then cooled down to 750 °C; the temperature was then raised to 800 °C, cooled to 600 °C, increased again to 700 °C, cooled down to 600 °C, and finally annealed at 600 °C for 3 days. After annealing, the flux was separated using a centrifuge to yield large bulk single crystals of Mn$_3$Ir.

Sample devices for transport measurements were fabricated using a six-probe configuration. After cleaning the sample surface, Au wires were bonded to the samples with A/B epoxy silver paste (EPO-TEK Company), and the assembly was cured at 150 °C for 5 minutes, ensuring robust electrical contacts. To further improve ohmic contact, we applied pulsed current to reduce the contact resistance, achieving final values of around 2 Ω. We performed longitudinal magnetoresistance and Hall resistance measurements using a Physical Property Measurement System (PPMS, liquid helium-free, Quantum Design, USA). To minimize the influence of thermal fluctuations on the weak AHE, each temperature point was stabilized for 1 hour before data acquisition. The measurement current was set at 3 mA. To obtain magnetoresistance data along the [111] direction, we adopted a non-conventional electrode layout (as shown in Figure **4a and 4d**). While this configuration introduces a substantial error in the absolute resistivity values, it provides reliable qualitative trends for evaluating the temperature and field dependence of resistivity and magnetoresistance. Magnetic measurements were carried out via a Magnetic Property Measurement System (MPMS, Quantum Design, USA). For all measurements, each temperature point was allowed to fully reach thermal equilibrium before collecting data. the authors thank the Center for Fundamental and Interdisciplinary Sciences of Southeast University for the support in Hall measurement/fabrication.

The elemental compositions of the Mn$_3$Ir samples were verified by SEM/EDS analysis. To obtain a reliable Mn$_3$Ir atomic ratio, the electron beam was focused on the polished cross-section of the sample, and more than fourteen points were collected for averaging (results shown in the Figure **S1**). The final composition was determined as Mn:Ir = 3:1, within the accuracy of the equipment. To identify the crystallographic orientation of the Mn$_3$Ir single crystals, we used a high-precision goniometer (angular error of 0.2°) and successfully obtained diffraction patterns along the [111], [110], and [100] directions

(as shown in the Figure **S2**).

We performed spin-polarized non-colinear density functional theory calculations using the Vienna ab initio simulation (VASP) package,[44, 45] with a generalized gradient approximation (GGA) using the Perdew-Burke-Ernzerhof exchange-correlation functional including spin-orbit coupling (SOC).[46] The relaxed lattice constant of 3.7103 Å, and a Γ-centered 15 ×15 × 15 k mesh sampling for the Brillouin zone of Mn$_3$Ir with a kinetic energy cutoff of 450 eV were used.

**Acknowledgements:** We sincerely thank Song Bao and Zihang Songbo from Prof. Jinsheng Wen's group at Nanjing University for providing support with the Laue diffraction measurements.

**Funding:** This work was supported by the National Key Research and Development Program of China (2023YFA1406600), the National Natural Science Foundation of China (Grant number 12274068), the open research fund of Key Laboratory of Quantum Materials and Devices (Southeast University), Ministry of Education, the National Key Research and Development Program of China (2022YFA1503103), Natural Science Foundation of Jiangsu Province, Major Project (BK20222007), the Fundamental Research Funds for the Central Universities (grant numbers 3207022201A3, 4060692201/006) and High-level personnel project of Jiangsu Province. The authors acknowledge the computational resources from the Big Data Center of Southeast University and the National Supercomputing Center of Tianjin. Support for this program/project/study is provided by RIXS endstation at BL09Uc in SSRF.

**Author contributions:** X.G., R.W., and B.Z. contributed equally to this work. T.C. conceived the project. T.C. planned the experiments. T.C., X.G., R.W. and H.W. grew the Mn$_3$Ir single-crystals. T.C., and X.G. prepared samples. X.G., R.W., and B.Z. carried out the transport and magnetization measurements and analyzed the data. R.W., X.G. and T.C. performed SEM/EDS and Laue measurments. S.Y. carryout DFT calculation. X.G. and T.C. prepared the Figures. T.C. wrote the paper. All authors discussed the results and commented on the manuscript.

**Conflict of Interest:** The authors declare no conflict of interest

**Captions and Figures**

**Figure 1.** presents the cleavage energy of Mn$_3$Ir for various crystallographic planes under distinct magnetic configurations, calculated via DFT method. a) shows the non-collinear triangular AFM order of Mn atoms in the reported L1$_2$-structured Mn$_3$Ir (gray spheres: Ir; purple spheres: Mn; red arrows: Mn magnetic moments), and the hypothetical FM order where all Mn spins align along [001]. b) quantifies the cleavage energies along <001>, <110>, and <111> directions after full structural and magnetic relaxation, revealing a minimal energy difference (~0.02 eV/Å$^2$) for the (111) plane in the AFM state.

**Figure 2.** shows the experimental growth and structural characterization of a bulk single crystal of Mn$_3$Ir. a) presents the zigzag cooling curve used for crystal growth, where the composition ratios of Bi, Mn, and Ir were adjusted to grow single crystals of Mn$_3$Ir via a high-throughput method with Bi as the flux.  b) displays the X-ray diffraction (XRD) pattern of as-grown Mn$_3$Ir. The inset shows the zoomed-in view of RRD patterns, with red inverted triangles indicating trace Bi impurities. c), d) and e) demonstrate the (110), (100), and (111) Laue spots.

**Figure 3.** shows the magnetic properties of the bulk single-crystal Mn$_3$Ir. a) shows the temperature-dependent out-of-plane magnetization with the inset highlighting low-field hysteresis (M-H) behavior. b) temperature-dependent in-plane magnetization under equivalent thermal conditions, where the inset emphasizes low-field M-H response. c) displays the zero-field residential magnetization for in-plane and out-of-plane directions. d) shows M-T curves for $\mu_0H \perp [111]$ and $\mu_0H \parallel [111]$ under 1.5 T, where a distinct kink at 150 K is also viewed for $\mu_0H \perp [111]$.

**Figure 4.** shows the magnetic transport results of the bulk Mn$_3$Ir single crystal in the in-plane (111) and out-of-plane configurations. a) presents the in-plane (111) resistivity-temperature (RT) curve of the Mn$_3$Ir bulk single crystal, while d) displays the corresponding out-of-plane RT curve. The insets in both figures illustrate the six-probe measurement configuration. b), c) and e) compare the AHE for in-plane (111) and out-of-plane configurations across multiple temperatures, with insets of (b) and (c) depicting the Hall resistivity after subtracting background. Specifically, (b) contrasts the AHE under 9 T field cooling, (c) under zero-field cooling. f) displays the longitudinal MR for in-plane (111) and out-of-plane geometries after 9 T field cooling.

**Figure 5.** presents the AHC for in-plane and out-of-plane configurations, and the possible AFM arrangement in bulk Mn$_3$Ir single crystals. a) shows the temperature-dependent AHC between the in-plane (111) and out-of-plane configurations of Mn$_3$Ir. b) shows the AHC predicted by Hua Chen et al., under A (a concentric circle that is black inside and gray outside) and B a concentric circle that is gray inside and black outside) magnetic phases, and experimental data (color configurations correspond to Figure 5a) in the (111) plane, where the inset shows A, B magnetic phases and alternating ABA stacking magnetic phases along the [111] axis.

**Figure 1**

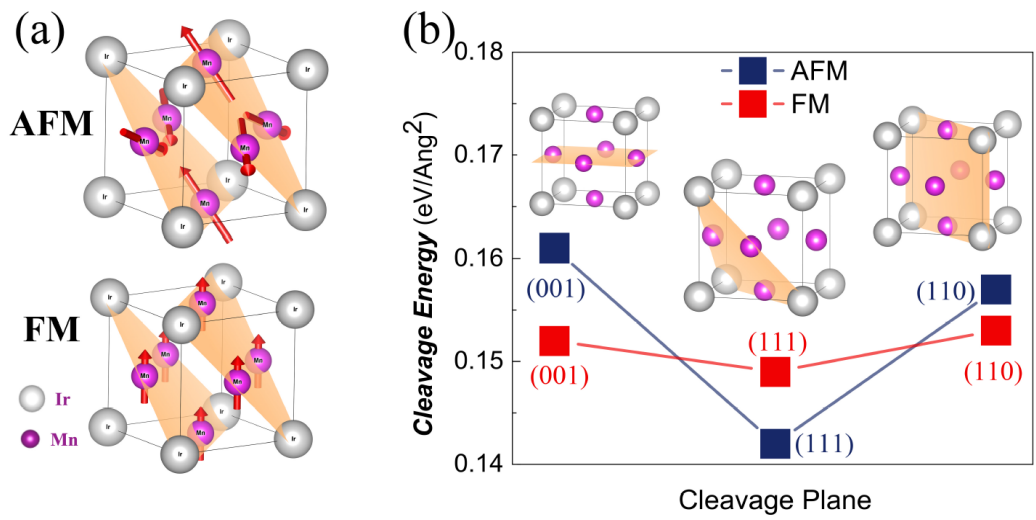

*Plotted by S.J. Yuan et al.*

**Figure 2**

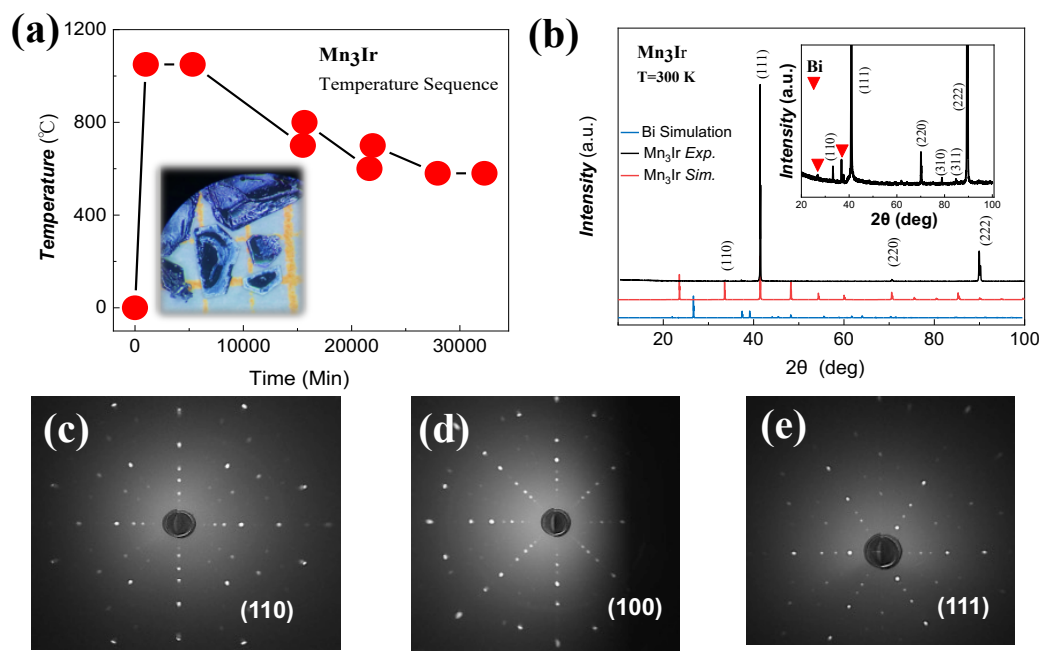

*Plotted by X. Gu et al.*

**Figure 3**

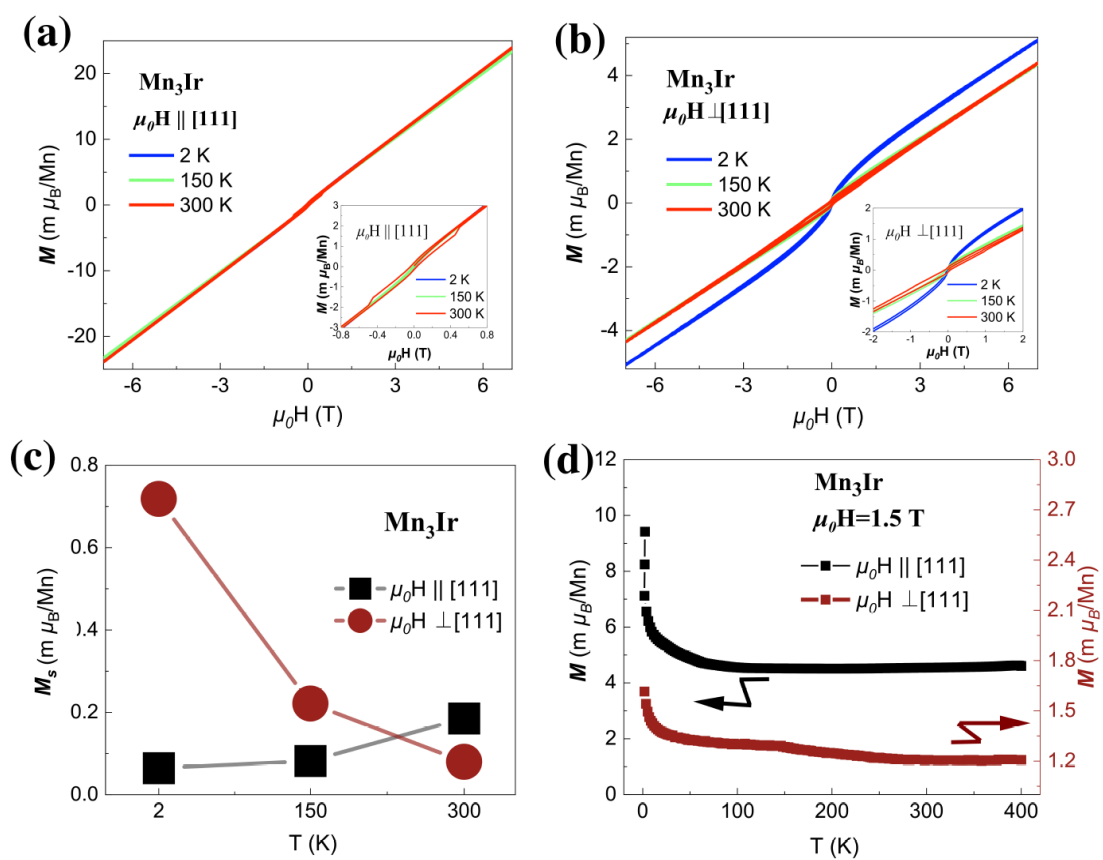

*Plotted by X. Gu et al.*

**Figure 4**

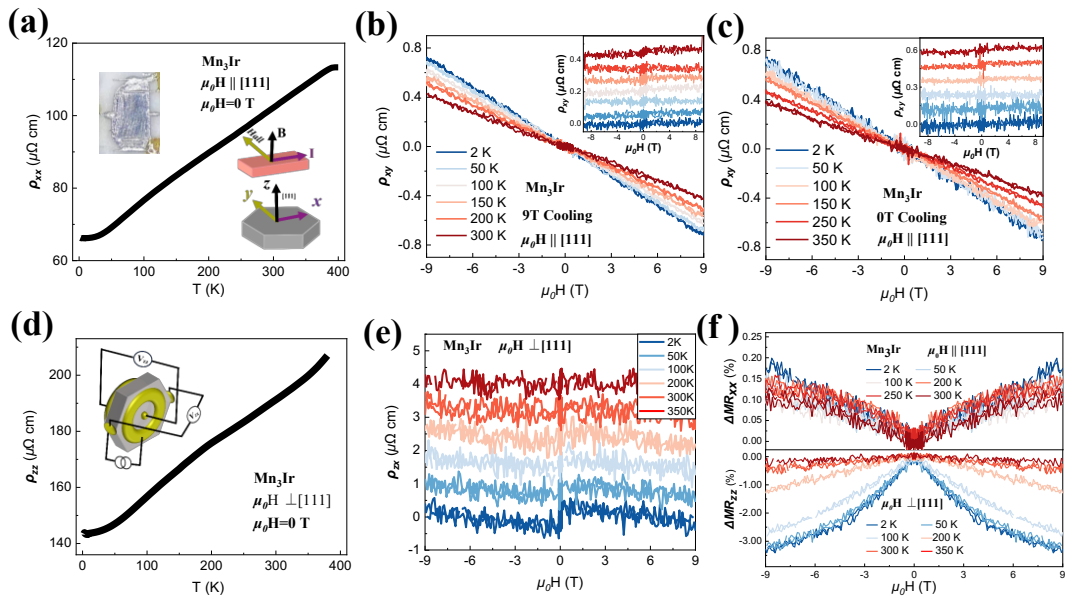

*Plotted by X. Gu et al.*

**Figure 5**

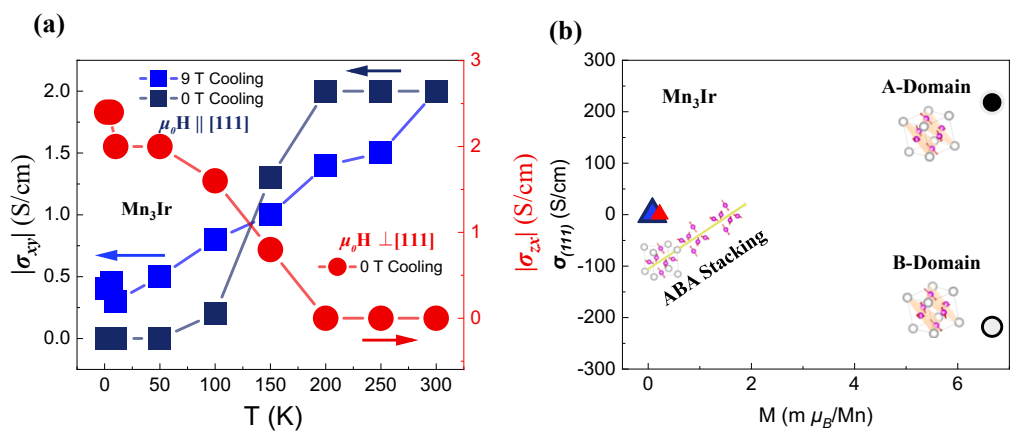

*Plotted by X. Gu et al.*

# Supplementary for "Zero-field Anomalous Hall Effect in Bulk Single Crystal Mn₃Ir"

Xin Gu, Ruoqi Wang, Bo Zhao, Haofu Wen, Kunquan Hong, Shijun Yuan[1,2]*, Taishi Chen[1,2]*, Jinlan Wang

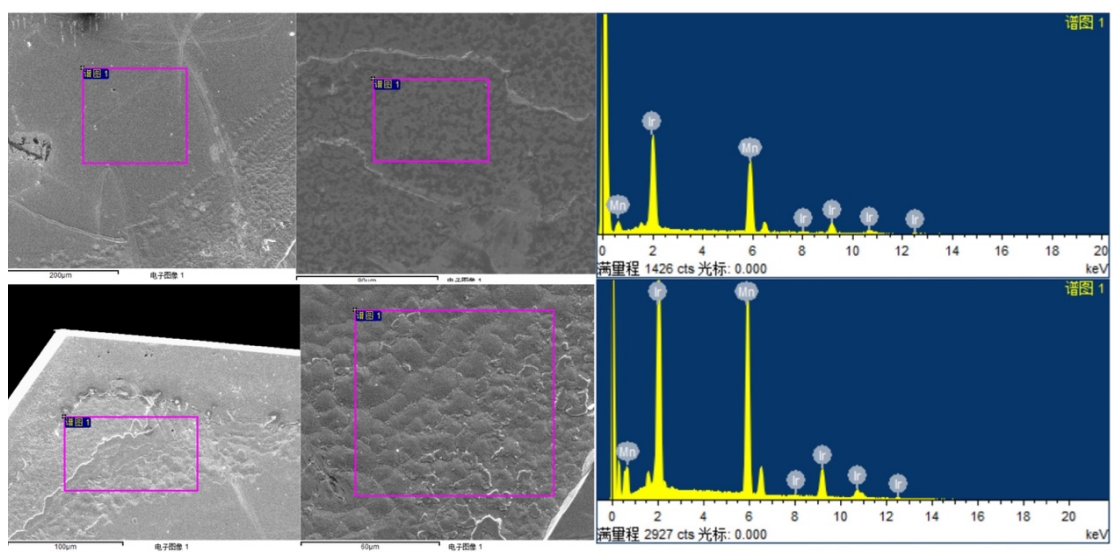

| Mn | Ir | Mn | Ir |
|---|---|---|---|
| 72.96 | 27.04 | 3 | 1.1118421 |
| 73.74 | 26.26 | 3 | 1.0683483 |
| 76.6 | 23.4 | 3 | 0.9164491 |
| 73.45 | 26.55 | 3 | 1.0844112 |
| 74.59 | 25.41 | 3 | 1.0219869 |
| 73.67 | 26.33 | 3 | 1.0722139 |
| 74.98 | 25.02 | 3 | 1.001067 |
| 73.45 | 26.55 | 3 | 1.0844112 |
| 73.51 | 26.49 | 3 | 1.0810774 |
| 73.54 | 26.46 | 3 | 1.0794126 |
| 73.41 | 26.59 | 3 | 1.0866367 |
| 75.07 | 24.93 | 3 | 0.9962701 |
| 75.7 | 24.3 | 3 | 0.9630119 |
| 73.74 | 26.26 | 3 | 1.0683483 |
| | Average=3:1.04 | | |

**Figure S1:** SEM/EDS results from the polished Mn₃Ir bulks, and below is the atomic ratio of Mn and Ir, which yields 3:1.

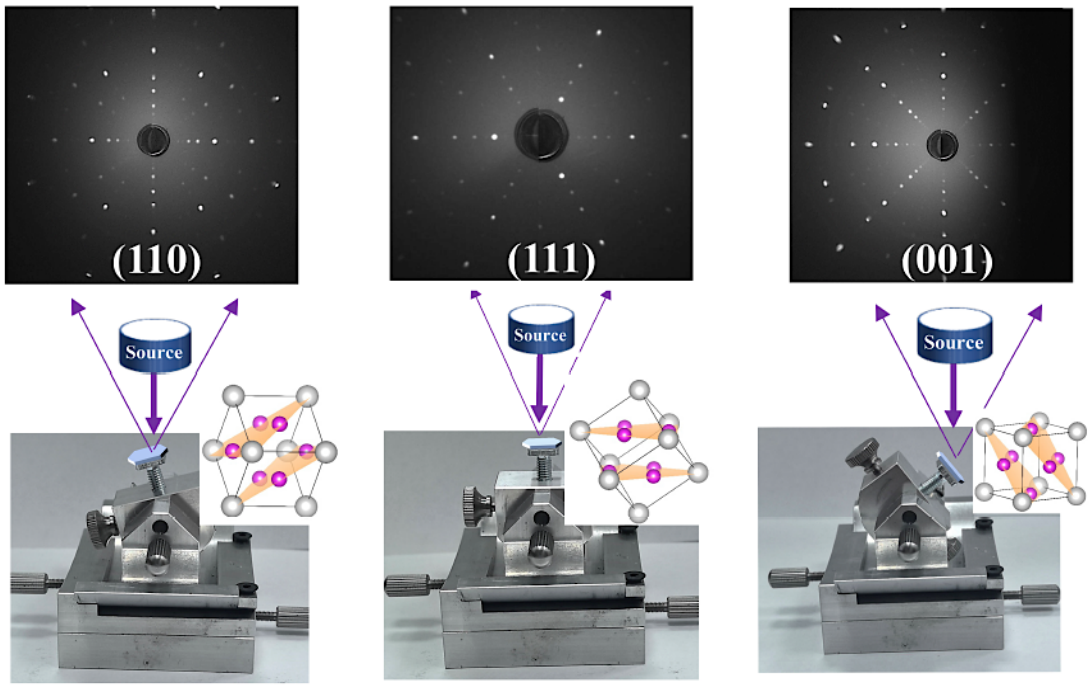

**Figure S2:** Laue patterns and measurement configurations. The orange triangles represent the related position of the Mn$_3$Ir samples in the primitive cell.

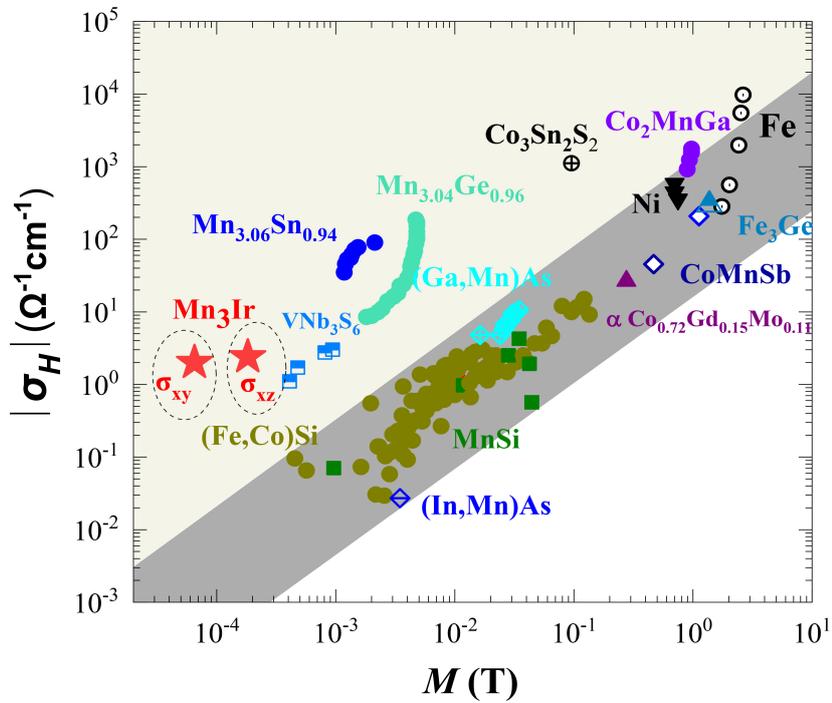

**Figure S3**: Scaling plot of anomalous Hall conductivity vs. magnetization. The black-shaded region represents the anomalous Hall effect arising from extrinsic mechanisms, while the light-yellow region corresponds to the intrinsic anomalous Hall effect originating from the Berry curvature.

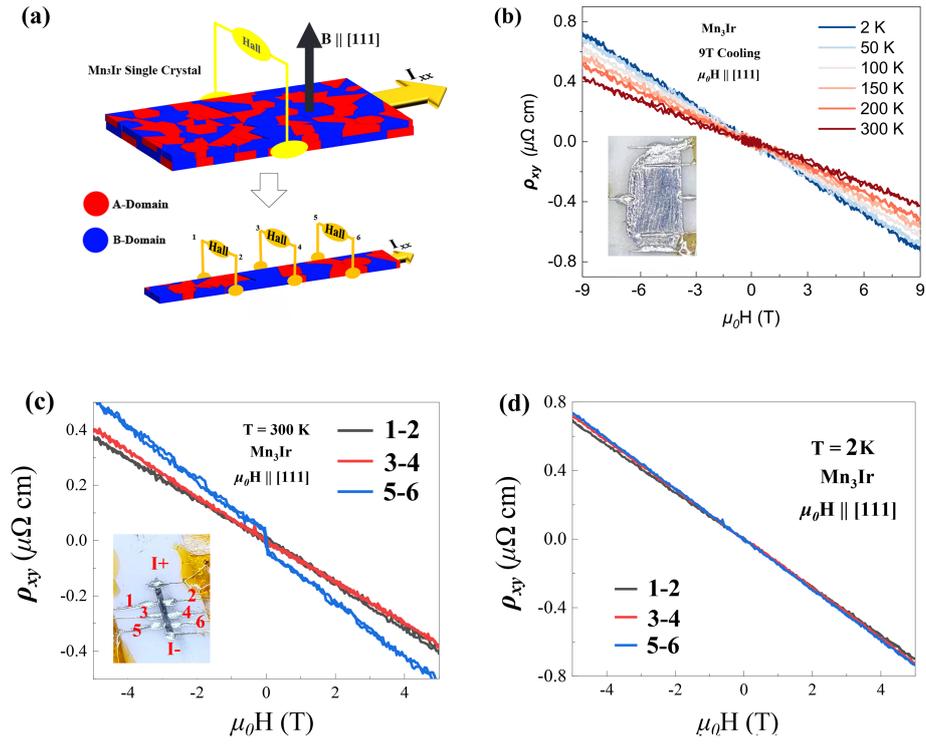

**Figure S4**: Enhanced AHE in miniaturized Mn$_3$Ir bulk crystals. a) In large-sized Mn$_3$Ir single-crystal bulks, the equal populations of A/B domains cancel the AHE. In contrast, small-sized samples can exhibit an imbalance in the number of A/B domains, leading to an enhanced anomalous Hall effect. b) AHE of large-sized Mn$_3$Ir bulk crystals measured at different temperatures: only a very small AHE is observed at 300 K (as shown in the main text). c) and d) AHE measured at different temperatures for a small-sized sample (2.2 mm × 130 $\mu$m × 30 $\mu$m) with three sets of Hall contacts. Notably, Hall contacts 5–6 show a pronounced enhancement in the AHE.